\documentclass[english,aps,prl,onecolumn, superscriptaddress,showpacs, amsmath]{revtex4-1}
\usepackage{graphicx}
\usepackage[colorlinks=true,linkcolor=blue,citecolor=blue]{hyperref}%

\begin{document}

	\title{Equilibration of quantum hall edge states and its conductance fluctuations in graphene p-n junctions}%
	
	\author{Chandan Kumar}
	\author{Manabendra Kuiri}
	\author{Anindya Das}
	\email{anindya@physics.iisc.ernet.in}
	\affiliation{Department of Physics, Indian Institute of Science, Bangalore 560 012, India}
	
	\begin{abstract}
		We report an observation of conductance fluctuations (CFs) in the bipolar regime of quantum hall (QH) plateaus in graphene (p-n-p/n-p-n) devices. 
		The CFs in the bipolar regime are shown to decrease with increasing bias and temperature. At high temperature (above 7 K) the CFs vanishes completely and the flat quantized plateaus are recovered in the bipolar regime. The values of QH plateaus are in theoretical agreement based on full equilibration of chiral channels at the p-n junction. The amplitude of CFs for different filling factors follows a trend predicted by the random matrix theory. Although, there are mismatch in the values of CFs between the experiment and theory but at higher filling factors the experimental values become closer to the theoretical prediction. The suppression of CFs and its dependence has been understood in terms of time dependent disorders present at the p-n junctions.
    \end{abstract}


\maketitle

\section{Introduction}
Graphene, with unique linear dispersion, exhibits anomalous Quantum Hall effect (QHE) because the number of chiral edge modes at the boundary of a graphene flake increases by multiple of four plus two\cite{zhang2005experimental,novoselov2005two,gusynin2005unconventional,neto2009electronic}. This results in half integer quantum Hall conductance plateaus as 4(n+1/2)$e^{2}$/h with integer number n\cite{zhang2005experimental,novoselov2005two,gusynin2005unconventional,neto2009electronic}. Even though each Landau level (LL) has four degeneracy coming from two spins and two valleys (K and K$'$), the half or anomalous effect can be explained by taking into account that the valley degeneracy gets lifted at the boundary of the graphene flake\cite{brey2006edge}. As a result for n = 0 LL there are only two chiral edge states coming from two spins degrees of freedom. The chiral edge states may not only occur at the boundary of a sample but can also form inside the sample. This can be realized by making a p-n junction in graphene\cite{abanin2007quantized,williams2007quantum,ozyilmaz2007electronic}, where the four chiral states propagate along the junction preserving  the valley degeneracy. 
\begin{center}
\begin{figure}[h]
\includegraphics[width=0.48\textwidth]{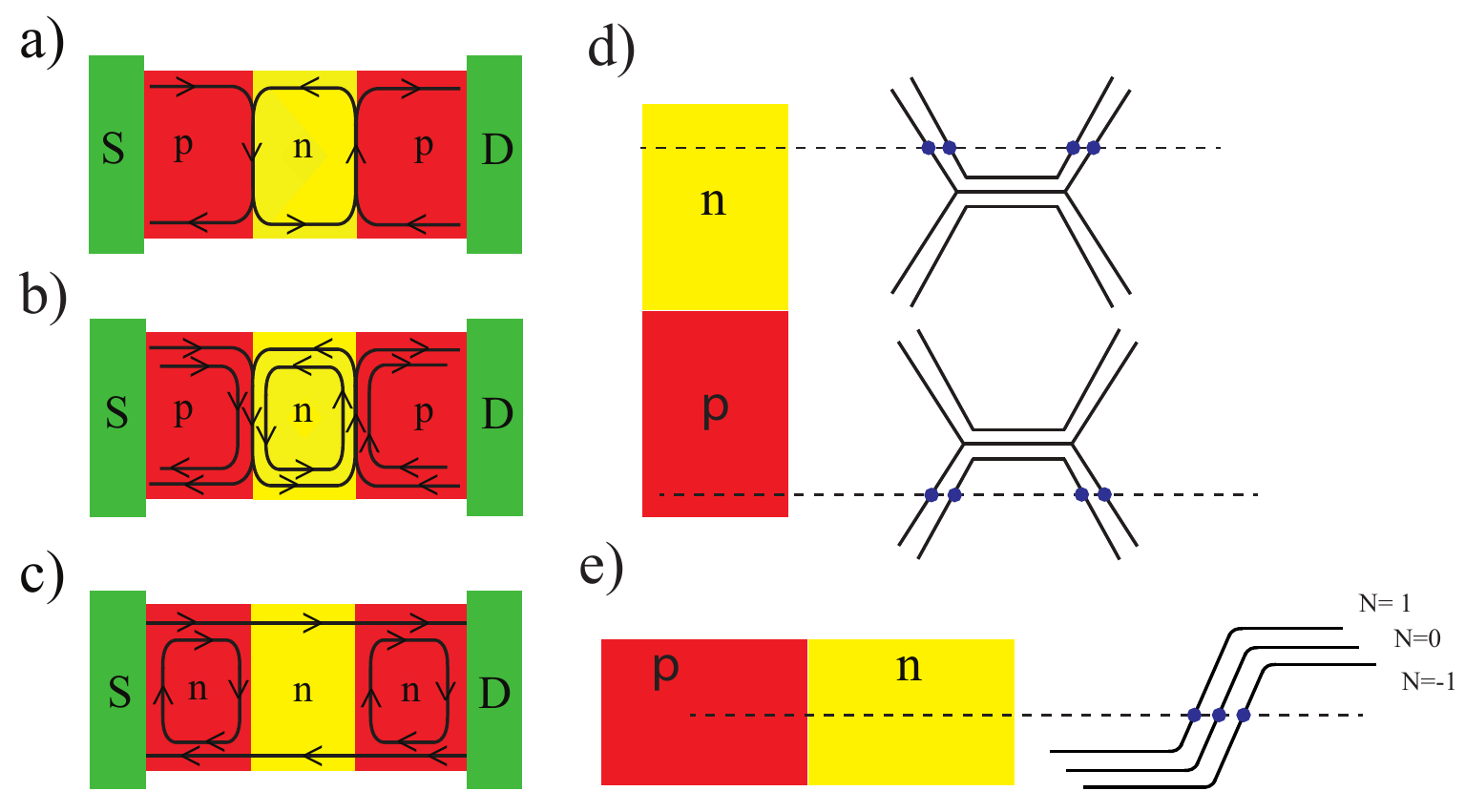}
 \caption{(color online) Schematics of QH edge states in a graphene based p-n-p device, where n and p region has filling factor of (a) 2,-2 and (b) 6,-6, respectively. Fig. (c) Shows the chiral edge states in unipolar regime, where the back gate filling factor ($\nu_{BG}$) is higher than the top gate filling factor ($\nu_{TG}$). Fig. (d) and (e) show the Landau level band diagrams at the boundary of the sample and the interface of a p-n junction for filling factor 6,-6.}
 \label{fig:images}
\end{figure}
\end{center}

\begin{figure*}[t]
\centering
\includegraphics[width=0.7\textwidth]{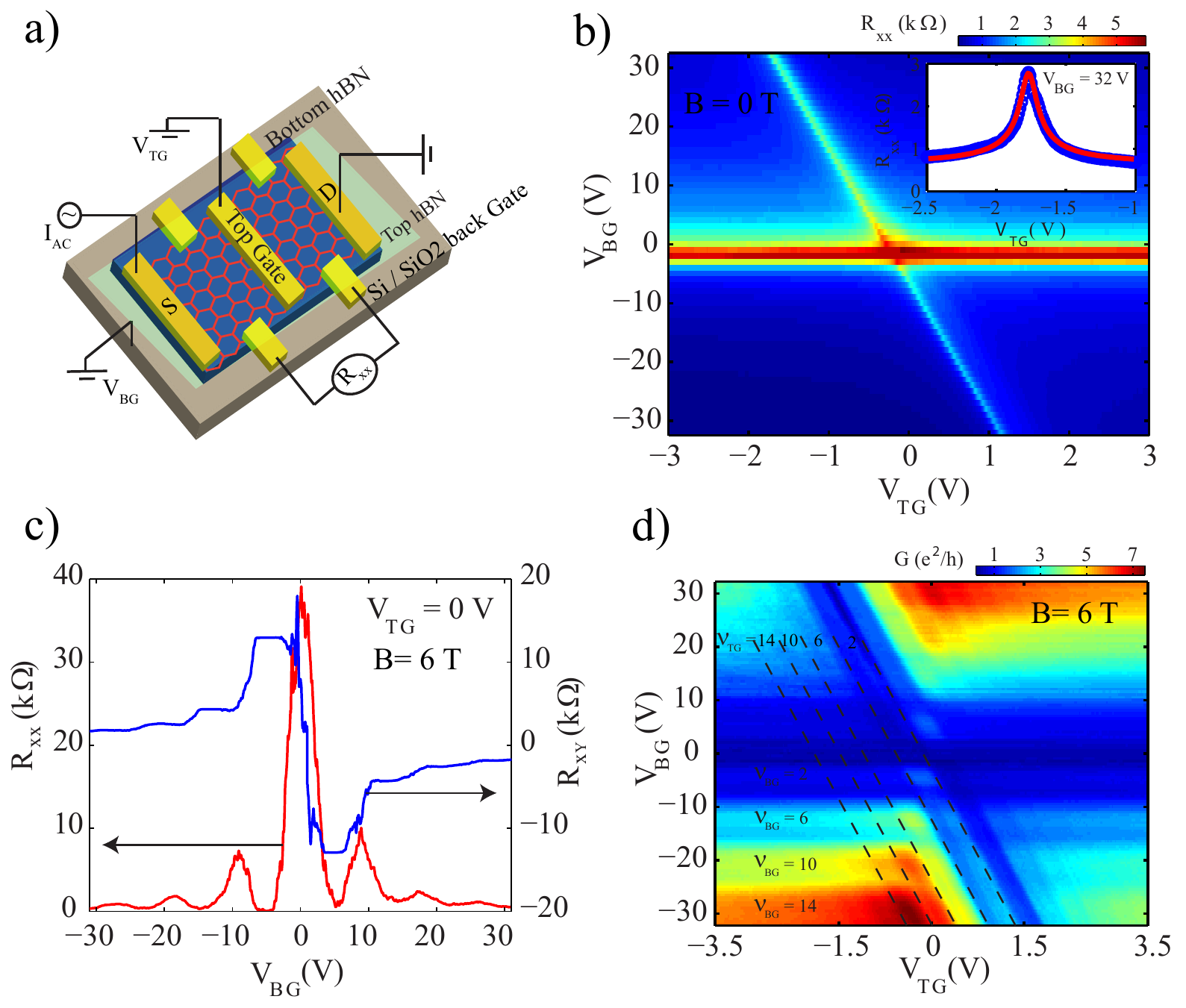}
 \caption{(color online) (a) Shows the device schematic, in which 300 nm thick Si$O_2$ acts as a global back gate and top thin hBN acts as a local top gate. (b) Shows the color plot of  two probe QH conductance as a function of $V_{BG}$ and $V_{TG}$ at 240 mK temperature. The inset of Fig. 2(b) Shows the cut line at $V_{BG}$= 32 V and the red curve shows the fit with equation (1). (c) Shows hall resistance ($R_{xy}$) and the longitudinal resistance ($R_{xx}$) of the device at 6 T perpendicular magnetic field and 240 mK temperature at $V_{TG}= 0$ V. The QH effect is shown by $R_{xy}$ plateaus and the vanishing of $R_{xx}$ at the corresponding $V_{BG}$. (d) Shows color plot of two probe QH conductance as
a function of $V_{BG}$ and $V_{TG}$ at 6 T perpendicular magnetic field and 240 mK temperature. The horizontal strips corresponds to the back gate filling factor ($\nu_{BG}$) while the diagonal line represents the top gate filling factor ($\nu_{TG}$).}
 
 \label{fig:electrical}
\end{figure*}
In recent years graphene p-n junction with perpendicular magnetic field has gained a lot of attention in condensed matter  physics\cite{williams2011snake,schmidt2013mixing,amet2014selective,rickhaus2015snake,taychatanapat2015conductance,kumada2015shot,matsuo2015edge,morikawa2015edge,
klimov2015edge,dubey2016tuning,fraessdorf2016graphene}. Such a p-n junction exhibits unprecedented phenomena like snake states\cite{williams2011snake,rickhaus2015snake,taychatanapat2015conductance}, where the interface state in a semi-classical picture alternatively propagate in the p and n side of the junction. However, in a high magnetic field one needs to consider the quantum picture of the chiral states. Graphene flake with p-n junction has $\lq$edge states' at the boundary as well as the $\lq$interface states' at the junction.
The conductance of such a p-n junction depends on how the electron coming from the source edge states enters into the interface states and flow out from the interface states to the counter propagating edge states 
This has been shown schematically in Figure 1 for a p-n junction having filling factor 2,-2 (Fig. 1a) and 6,-6 (Fig. 1b). Figure 1(c) and (d) shows the schematic of LL at the edge of graphene device and at the p-n junction interface respectively. The average conductance plateaus of such bipolar junctions (Fig. 1a and 1b) does not exhibit half integer values rather show fractional values as predicted by Abanin et.al~\cite{abanin2007quantized}, which depends on the equilibration of the interface states by disorders. 

The theoretical prediction based on random matrix theory by Abanin et.al \cite{abanin2007quantized} and numerical simulations by J. Li et.al
\cite{li2008disorder} predict that there will be large mesoscopic fluctuations in bipolar regime. Thus, for a given disorder configuration one can not observe the flat conductance plateaus. Theoretically, the flat conductance plateaus in bipolar regime were obtained by taking average over large number of disorder configurations or ensemble average. In an experiment, as there will be only one unique disorder configuration and thus, the CFs should emerge. For last one decade several experiments\cite{williams2007quantum,ozyilmaz2007electronic,
liu2008fabrication,lohmann2009four,velasco2009electrical,ki2009quantum,ki2010dependence,lohmann2009four,
velasco2010probing,jing2010quantum,woszczyna2011graphene,nam2011ballistic,velasco2012quantum,bhat2012dual,
schmidt2013mixing,amet2014selective,klimov2015edge} have been performed on graphene p-n junction devices. Most of the experiments were carried out on Si$O_2$ substrate as global back gate and Al$_{2}$O$_{3}$/ HSQ/ PMMA/ air bridge as a local top gate. They have reported the observation of fractional conductance plateaus in bipolar regime. However, the inevitable CFs were not observed. The surprising results were explained by considering the time dependent fluctuations of disorders\cite{abanin2007quantized} at the p-n junction. As a consequence the system transforms into an ensemble average quantity and exhibits flat conductance plateaus. Abanin et.al \cite{abanin2007quantized} suggested that the suppression of CFs could be also explained by the de-phasing mechanism originating from the coupling between the localized states in the bulk with disorder state at the p-n junction.\\

In this paper we have carried out Quantum Hall measurements in hBN protected graphene devices. Our data for the first time reveals the evidence of CFs superimposed on QH plateaus in the bipolar region (p-n-p/n-p-n) of graphene. The CFs are shown to decrease with bias voltages as well as with temperature beyond a critical value. The CFs vanish completely at high temperature (7 K) and the exact value of the QH plateau was recovered (Ref\cite{dubey2016tuning}). However, there is a discrepancy between the amplitude of CFs of our experiment with the theoretical values (Abanin et al.\cite{abanin2007quantized}) at low filling factors. Interestingly, at higher filling factors the experimental values become very close to the theoretical prediction. In order to understand the suppression of CFs at lower filling factors we have also carried out $1/f$ noise measurement in the bipolar region. The above measurements suggest the existence of time dependent disorder in the devices.

\section{Device fabrication}
The schematic of the device is shown in figure 2(a). First a thin layer of hBN ($\sim$ 20 nm) was transferred on a 300 nm thick Si$O_2$ , which was followed by transfer of a graphene using dry transfer technique~\cite{zomer2011transfer}. The contacts were fabricated using standard electron-beam lithography technique, followed by Cr/Au (5 nm/ 70 nm) deposition. The prepared heterostructure was vacuum annealed at 300$^{\circ}$C for 3 hours which was followed by a thin hBN ($\sim$ 13 nm) transfer on the prepared Graphene-hBN-Si$O_2$ heterostructure. Finally, the top gate was defined using lithography, followed by Cr/Au (5nm/70 nm) deposition. The 300 nm thick Si$O_2$ serves as global back gate and thin top hBN ($\sim$ 13 nm) as local top gate. Different combination of back and top gate voltages leads to the formation of unipolar (p-p-p/n-n-n) and bipolar (n-p-n/p-n-p) region. All the measurements were performed using standard lock-in technique at 6 T perpendicular magnetic field (except Fig. 2b) at a base temperature of 240 mk in a He3 refrigerator. 
\section{Results and discussion }
Figure 2(b) shows the 2D color plot of resistance as a function of back gate ($V_{BG}$) and top gate ($V_{TG}$) voltages at 240 mK at zero magnetic field. From the slope of diagonal line we calculate top hBN thickness to be 13 nm, which implies that the relative coupling of top gate
with back gate is $\sim 23$. The inset of Figure 2(b) shows the cut line at $V_{BG}$= 32 V. Red line is a fit to the $R-V_{TG}$ curve with the following equation\cite{venugopal2011effective}:

\begin{equation}
 R=R_{c}+\frac{L_{TG}}{we\mu\sqrt{\delta n^2+n^2}}
\end{equation}

where $R_{c}$, $\mu$, $L_{TG}$ and $w$ are the contact resistance, mobility, length and width of top gate region, respectively. From the fitting we obtained a mobility of 25000 $cm^2/Vs$ and contact resistance of 500 $\Omega$, with $L_{TG}= 1.8$ $\mu m$ and $w= 1.9$ $\mu m $.
\begin{figure}[h]
\includegraphics[width=0.51\textwidth]{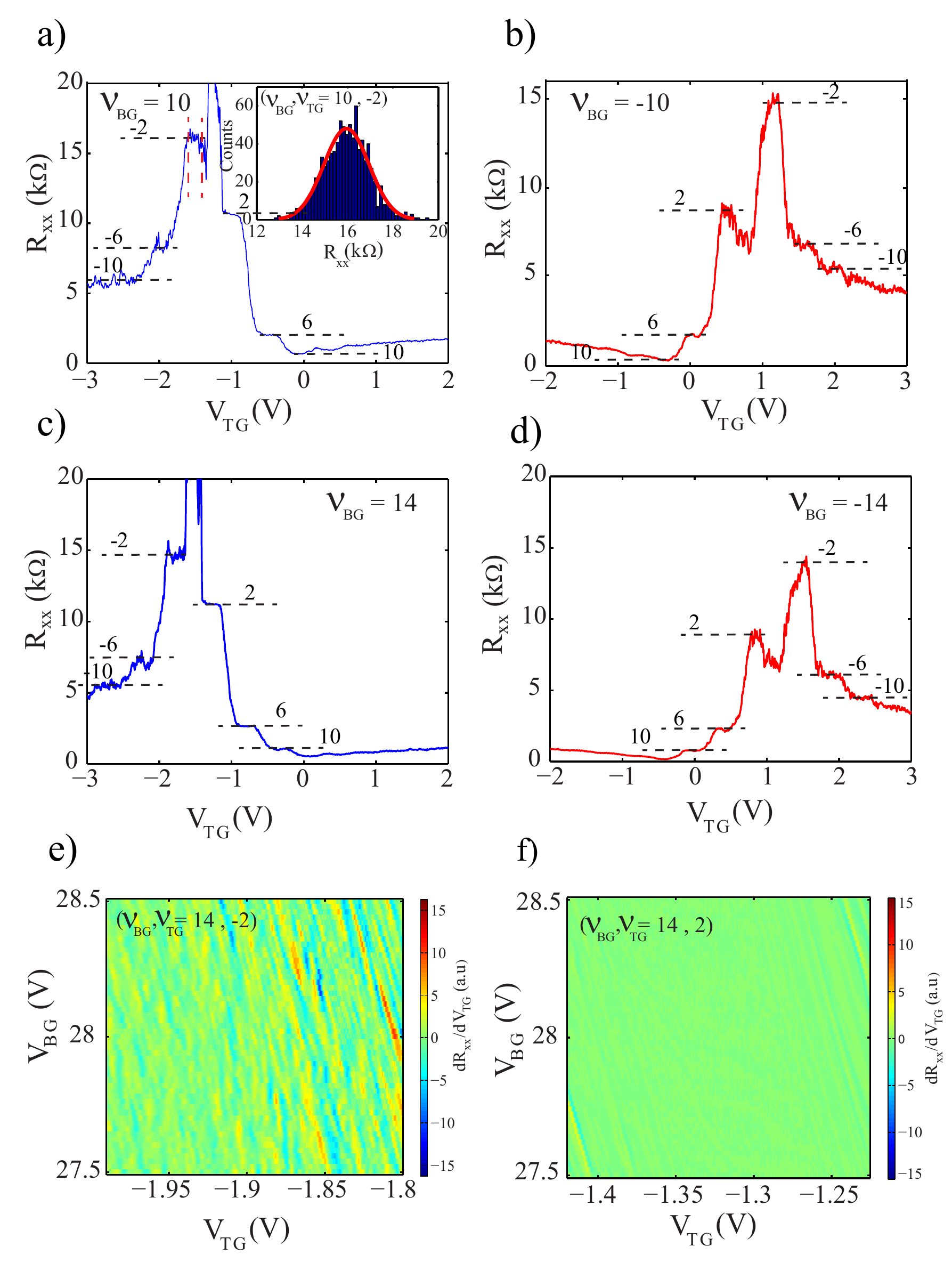}
 \caption{(color online) Figure (a) (b) (c) (d) shows longitudinal resistance, $R_{xx}$ as a function of $V_{TG}$ for $\nu_{BG}= 10, -10 , 14 ,-14$ respectively. The horizontal dashed line shows the different top gate filling factor. The inset of (a) shows the distribution of fluctuation for 10,-2 QH plateau and is of pure gaussian nature. The data used for histogram is marked by red vertical dashed line in Fig 3(a).  (e) and (f) Show color plot of transresistance as a function of $V_{BG}$ and $V_{TG}$ for
 14,-2 (bipolar) and 14,2 (unipolar) QH plateau, respectively.}
 \label{fig:optical}
\end{figure}

\subsection{Quantum hall plateaus in unipolar and bipolar regime}
Figure 2(c) shows longitudinal resistance ($R_{xx}$) and hall resistance ($R_{xy}$) measured at 6 T perpendicular magnetic field at $V_{TG}= 0 $ V. We see the clear LL plateaus at $\nu_{BG} = \pm 14, \pm 10, \pm 6,\pm 2$ and the vanishing of $R_{xx}$ components at the corresponding back gate voltages. The obtained QH resistance values are in exact agreement with the theoretical value for single layer graphene.\\

The resistance of a graphene device in high magnetic field depends on the filling factors $\nu_{BG}$, $\nu_{TG}$
\cite{williams2007quantum,ozyilmaz2007electronic,abanin2007quantized,dubey2016tuning}. In the unipolar regime (with
 $\nu_{BG} > \nu_{TG}$) the QH edge states in the top gate region is fully transmitted as shown in Figure 1 (c) and the resistance is given as\cite{dubey2016tuning}
\begin{align*}
  R_{2Probe}&=\frac{h}{e^2} |\nu_{TG}|, & R_{xx}= \frac{h}{e^2} \frac{|\nu_{BG}|-|\nu_{TG}|}{|\nu_{BG}||\nu_{TG}|}
\end{align*}
 $R_{2probe}$ is the two probe resistance measured between the source and drain (S and D of Fig. 2a). For the bipolar regime, the average value of the resistance depends on the equilibration of the interface states (Fig. 1a and 1b) and in the case of full mixing the resistance can be written as\cite{dubey2016tuning}
\begin{align*}
R_{2Probe}&=\frac{h}{e^2} \frac{2|\nu_{TG}|+\nu_{BG}}{|\nu_{BG}||\nu_{TG}|}, & R_{xx}= \frac{h}{e^2} \frac{|\nu_{TG}|+|\nu_{BG}|}{|\nu_{BG}||\nu_{TG}|}
\end{align*} 
Figure 2(d) shows the color plot of two probe conductance measured at 
B = 6 T  as a function of $V_{BG}$ and $V_{TG}$. The horizontal strips and the dashed diagonal lines in Fig. 2d represents $\nu_{BG}$ and $\nu_{TG}$, respectively. We observe clear QH plateaus for $\nu_{BG}$, $\nu_{TG} = \pm 14, \pm 10, \pm 6,\pm 2$ in the unipolar regime.\\

Figure 3(a), (b), (c) and (d) shows $R_{xx}$ as a function of $V_{TG}$ for $\nu_{BG}$ = 10, -10, 14, -14, respectively. We obtain the expected quantized resistance values in the unipolar region. In the bipolar regime we observe the resistance fluctuation on top of the expected quantized quantum hall plateaus \cite{abanin2007quantized,williams2007quantum,ozyilmaz2007electronic,dubey2016tuning}. The dashed horizontal lines 
in Fig. 3 (a), (b), (c) and (d) indicates the expected quantized resistance values in the unipolar and the bipolar regime. The similar CFs in the bipolar regime for $\nu_{BG} = \pm 6$ has been shown in S.I.\\
 
In order to investigate the nature of CFs we have measured trans- resistance $dR_{xx}$/$dV_{TG}$ as a function of $V_{BG}$ and $V_{TG}$ for $\nu_{BG}, \nu_{TG} = 14,-2$ plateau and is shown as 2D color plot in Fig. 3e. Figure 4 (c) shows the auto correlation of the data shown in Fig. 3e. It can be clearly seen from Fig. 3e and 4c that there is no clear pattern in the resistance fluctuations, thus, ruling out the possibility of any kind of Fabry-Perot interference, snake state, Aharonov Bohm interference or the oscillations arising from penetration of magnetic flux in the insulating area between the co-propagating p and n quantum Hall edge channels\cite{morikawa2015edge}. The distribution of fluctuations show pure gaussian nature and has been shown in the inset of Fig. 3a. The histogram is extracted from the data shown in figure 3a, marked by dashed vertical red lines. Figure 3 (f) shows the 2D color plot of trans- resistance as a function of $V_{BG}$ and $V_{TG}$ for 14, 2 QH plateau (unipolar regime), indicating the absence of fluctuations in the unipolar regime.

\subsection{Conductance fluctuation}
The universal conductance fluctuation (UCF) is an ubiquitous quantum interference phenomena occurring in disordered mesoscopic devices at low temperatures. It has been a topic of great interest in graphene devices\cite{skakalova2009correlation,lee2012transconductance,rahman2014quantum1}. As the coherent electron travel in all possible paths it get scattered repeatedly and the interference of these coherent waves give rise to UCF. When the phase coherence length is larger or comparable to the sample size, the conductance
fluctuates with the universal magnitude of $e^2$/h and is independent of the degree of disorder or the geometry of the device.
It is predicted by Abanin et.al\citep{abanin2007quantized} that in a fully coherent regime conductance would exhibit UCF and the magnitude of UCF will depend on the number of channels as follows:
\begin{equation}
var(g)=\frac{\nu_{BG}^2 \nu_{TG}^2}{(|\nu_{BG}|+|\nu_{TG}|)^2 [(|\nu_{BG}|+|\nu_{TG}|)^2-1]}
\end{equation}
As mentioned earlier the resistance in our experiment was measured in a four-probe geometry. In order to compare with the theoretical predictions \cite{abanin2007quantized} we need to convert the resistance fluctuations into the conductance fluctuations. We have used the following method. The two probe conductance  can be written as 

$$ G_{2Probe}= \frac{1}{R+std (\Delta R_{xx})+ std (\Delta R_{xy})}= \frac{1}{R+2std(\Delta R_{xx})}$$ where, $R = R_{xx}+R_{xy}$ and we have assumed that $std (\Delta R_{xx})=std (\Delta R_{xy})$.
Thus the maximum change in conductance can be written as
\begin{align*}
G_{1}-G_{2}&= \frac{1}{R-2std (\Delta R_{xx})}-\frac{1}{R+2std (\Delta R_{xx})}\\
& \sim  \frac{4 std (\Delta R_{xx})}{R^2}
\end{align*}
Hence, the standard deviation of conductance fluctuation can be written as
\begin{align}
&std(\Delta G_{2Probe})= \frac{2std(\Delta R_{xx})}{(<R_{xx}>+<R_{xy}>)^2}
\end{align} 
 \begin{center}
\begin{figure}[h]
\includegraphics[width=.5\textwidth]{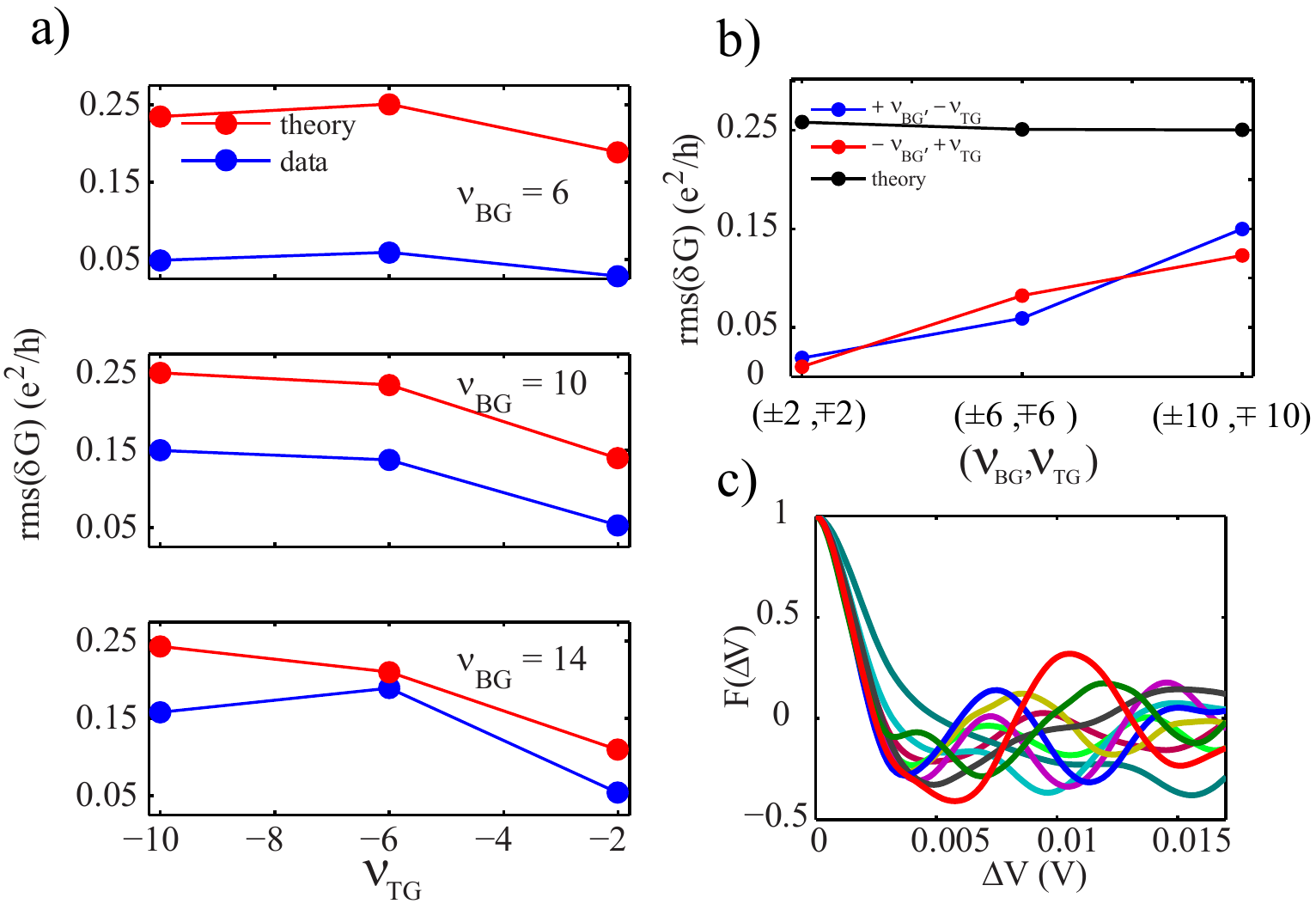}
 \caption{(color online)(a) Shows CF as a function of $\nu_{TG}$ for different set of $\nu_{BG}$ (b) CF as a function of $\nu_{BG}, \nu_{TG}$ (c) Shows the auto- correlation of the data shown in Fig. 3e. The different curves represents the different back gate voltages ranging from 27.5 V to 28.5 V with spacing of 0.1 V.}
 \label{fig:ionic gating}
\end{figure}
\end{center}
 Figure 4(a) shows the magnitude of CFs as a function of $\nu_{TG}$, for different set of $\nu_{BG}$ in the bipolar regime. We see that although the experimental and the theoretical values \cite{abanin2007quantized} have the same trend but there are large mismatch between the two values. Interestingly, with increasing $\nu_{BG}$, the discrepancy decreases and the values of experimental CFs approaches closer to the theoretical prediction. Figure 4(b) shows CFs as a function of  $\pm \nu_{BG}, \mp \nu_{TG}$ and the values are same for p-n-p or n-p-n configuration. It is evident that with increasing filling factor, the mismatch between the measured CFs and the theoretical value decreases.\\

\subsection{Dependence of conductance fluctuation on bias and temperature}
Next we tune to $\nu_{BG}$= 10 and study the resistance fluctuation with bias and temperature. 
 Figure 5 (a) shows the resistance fluctuation as a function of $V_{TG}$ at $V_{Bias}$ = 0 V and 1.6 mV. We notice that at $V_{Bias}$ = 1.6 mV the flat QH plateaus in the bipolar regime are almost recovered. Similarly, the resistance fluctuation was also killed with temperature. Figure 5b shows that at 9 K the fluctuation vanishes completely and we obtain the flat quantized plateaus in the bipolar regime. \\
 
\begin{center}
\begin{figure}[h]
\includegraphics[width=.515\textwidth]{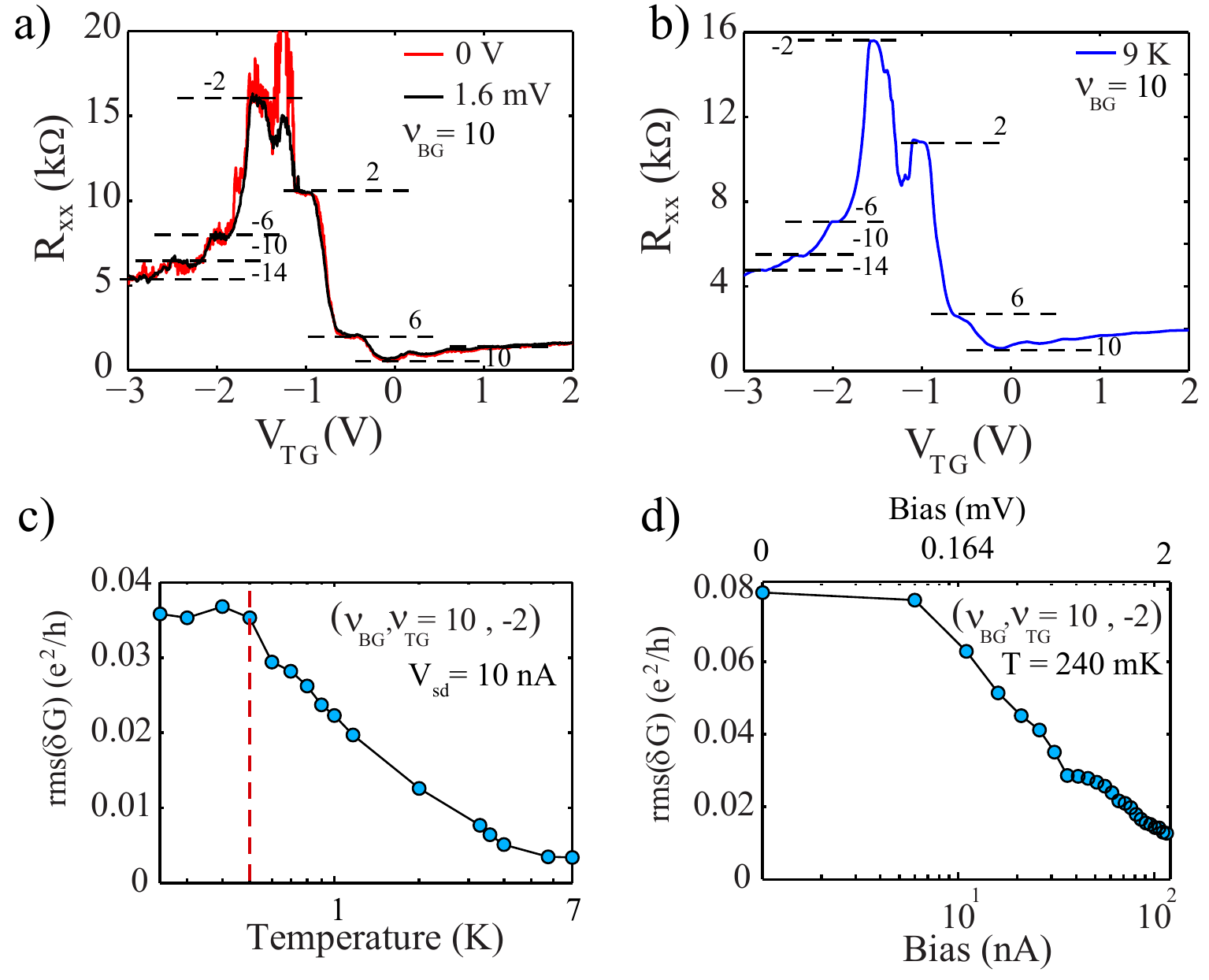}
 \caption{(color online) (a) Shows QH resistance fuctuation for $\nu_{BG}$ = 10 as a function of $V_{TG}$ at 0 bias (red line) and at 1.54 mV (black line). (b) shows $R_{xx}$ as a function of $V_{TG}$ for $\nu_{BG}$= 10 at 9 K. The dashed line shows the expected QH resistance value. Fig. (c) and (d) shows CF amplitude as a function of temperature and bias for 10,-2 QH plateau respectively.}
 \label{fig:bais}
\end{figure}
\end{center}

Figure 5c shows $\delta G$ as a function of temperature for 10,-2 QH plateau. We see that fluctuation amplitude remains constant till $\sim$ 400 mK ($T_{c}$) and starts to decrease beyond this temperature. There were no detectable CFs beyond T = 7 K. Figure 5 (d) shows $\delta G$ as a function of bias for 10,-2 QH plateau, obtained from figure 2 of S.I. It can be seen that $\delta G$ remains constant till 6 nA which is equivalent to $\sim$ 130 $\mu $eV ($V_{sd}^{c}$) and decreases beyond this point. The complete suppression of conductance flutuation is seen $\sim 2$ mV. The complete suppression of CFs at temperature, T = 7 K or at bias, $V_{sd}= $ 2 mV are of similar energy scale(2 mV $\sim 3.5 K_{B}T, T = 7 K$). From fig. 5c (fig. 5d) it is clear that above $T_{c}$ ($V_{sd}^{c}$) thermal energy is the main source of CFs suppresion. The energy scale can further be justified by comparing thermal length ($L_{T}$) versus the length of the p-n junction interface. The $ L_{T} = \sqrt{\frac{\hbar D}{k_{B} T}}$, where $\hbar$ is the planks constant, $k_{B}$ is the Boltzman constant, T is the temperature and  $D$ is the diffusion constant with a value of 250 $\times 10^{-3}$ $m^2/s$ ( see S.I). At T = 400 mK, $L_{T} \sim 1.6$ $\mu m $, which is comparable to the length of the p-n junction of our device. Although the expression used for thermal length is valid at zero magnetic field, but we find it to be in good agreement with our data.

\subsection{Conductance suppression mechanism}	
From Fig. 5c and 5d it is clear that above $T{c}$ and $V_{sd}^{c}$ the origin of CFs suppression is thermal energy broadening. However, the saturation value of CFs below $T_{c}$ and $V_{sd}^{c}$ are far from the predicted theoretical value\cite{abanin2007quantized}. Abanin et. al has predicted the following origins as the source of CFs suppression: \newline
(i) time dependent fluctuations of disorders at the p-n junction \newline
(ii) some intrinsic de-phasing mechanism due to the coupling between localized states in the bulk with disorder states at the pn interface\newline
A qualitative scheme to understand the suppression of CFs based on above origins has been depicted in Fig 6. It can be seen from Fig. 6a that the transmission at the p-n junction depends on how the discrete disorder energy levels are aligned with the Fermi energy position. Therefore, as a function of $E_{F}$ shifting, Fermi energy may align with one of the disorder levels or it may lie in between the disorder levels. For the case of alignment, half will be transmitted and half will be reflected back, in contrary everything will get reflected back for the non-alignment case, thus the CFs arises as a function of gate voltages or the Fermi energy shift.  However, when the energy levels of the disorders are broadened (Fig. 6b), the transmission will relatively less sensitive on the location of the Fermi energy position, as a result the CFs will be suppressed. The origin of the broadening may arise due to above mentioned two mechanisms. The first origin coming from the time dependent disorder has been investigated by 1/f noise measurement. \\
The $1/f$ noise measurement is a versatile tool to capture the time dependent phenomena like traping-detraping of charges, disorder scattering, mobility fluctuation etc. There have been extensive studies of $1/f$ noise measurement in graphene devices\cite{lin2008strong,pal2011microscopic,rahman2014quantum,kumar2016tunability}. The schematic of noise measurement technique with experimental details and the defination of noise (A) has been mentioned in the S.I.\\
Figure 6c shows the $1/f$ noise power spectral density as a function of frequency in both unipolar (2,2; 6,6; 10,10) and bipolar regime (2,-2; 6,-6; 10,-10). It can be seen that in the unipolar regime the $1/f$ noise magnitude is very small and comparable to the background noise level (horizontal black solid line in Fig. 6c). This is indeed expected in the QH regime because the transprt through the chiral edge states are ballistic in nature. The detectable $1/f$ noise in the unipolar regime may arise from the contacts. However, in the bipolar regime transport occurs by tunneling between the LL via the defect states located at the p-n junction. Thus, one can expect to have $1/f$ noise coming from the time dependent fluctuation of disorders in the bipolar regime. It can indeed be seen from Fig.6c that in the bipolar regime the magnitude of $1/f$ noise is one order higher compared to unipolar regime.
\begin{center}
\begin{figure}[h]
\includegraphics[width=.5\textwidth]{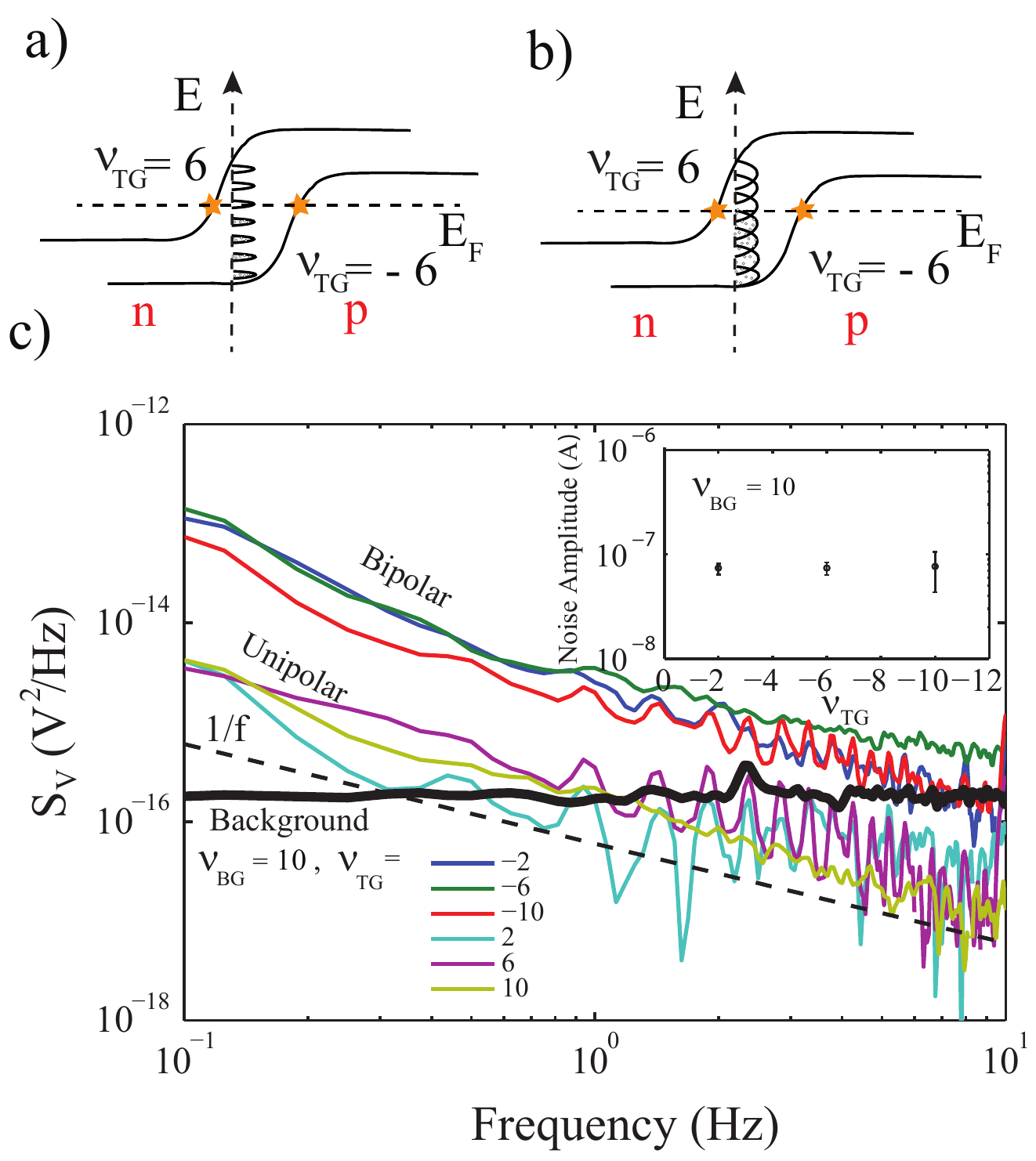}
\caption{(color online) (a) Schematic for conductance fluctuation as a function of gate voltage for (a) sharp disorder levels and disorder states are aligned with Fermi energy level (b) the broadend disorder energy level, where CFs does not depend appreciably on the Fermi energy location (c) Shows 1/f noise as a function of frequency in unipolar and the bipolar regime. The dashed line indicates the ideal 1/f noise behaviour. The inset shows the noise amplitude as a function of $\nu_{TG}$ for $\nu_{BG}$=10.}
 \label{fig:ionic gating}
\end{figure}
\end{center}
 The inset of Fig. 6c shows the noise amplitude with filling factor in the bipolar regime. It is seen that the magnitude of noise in bipolar regime remains almost constant with increasing filling factor. This implies that the fluctuation of disorder levels does not change appreciably with increasing filling factor. Hence, increment in $\delta G$ with increasing filling factor (Fig. 4a and 4b)can not be explained by the first origin. Thus, the time dependent fluctuation of disorder at the p-n junction is not sufficient enough to explain the suppresion of CFs. Hence we conjecture that the second dephasing mechanism is playing a significant role which can explain the increase in CFs with increasing filling factor. Further studies are required to understand the role of above mechanism in CFs suppression.
\section{Conclusion}
In conclusion we show first experimental signature of CFs in the bipolar regime in QH regime. The study also shows the crucial role of disorders in mixing the interface states at the p-n junction. By comparing the experimental values of CFs with the theoretical prediction as well as their dependence with increasing filling factors and $1/f$ noise measurements help us to separate out the contribution coming from the time dependent fluctuations of disorders at the p-n junction and de-phasing mechanism from the bulk. Our study will help in further understanding of edge state equilibration process at a p-n junction.\\
\section{Acknowledgements}
We thank Tanmoy Das, Jeil Jung, Nicolas Leconte, Adhip Agrawal and Vijay Shenoy for helpful discussion. A.D. thanks
nanomission under Department of Science and Technology, Government of India.
\section{Appendix A. Supplementary data}
Diffusion constant , 1/f noise measurement technique and experimental details, measurement performed on second device.

\newpage
\begin{center}
	\textbf{\large Supplementary Material: Equilibration of quantum hall edge states and its conductance fluctuations in graphene p-n junctions}
\end{center}
\begin{center}
	Chandan Kumar${}^{1}$, Manabendra Kuiri${}^{1}$, and Anindya Das${}^1$
	
	\it{${}^1$Department of Physics, Indian Institute of Science, Bangalore 560012, India}
\end{center}
\section{Conductance fluctuation in bipolar regime for different back gate filling factor}

Figure 7(a)(b) shows $R_{xx}$ as a function of $V_{TG}$ for $\nu_{BG}= 6,-6$ respectively. Clear fluctuation is visible in the bipolar regime on the quantum hall plateau while in the unipolar regime flat quantum hall plateau is obtained. 

\begin{center}
	\begin{figure}[h]
		\includegraphics[width=1.05\textwidth]{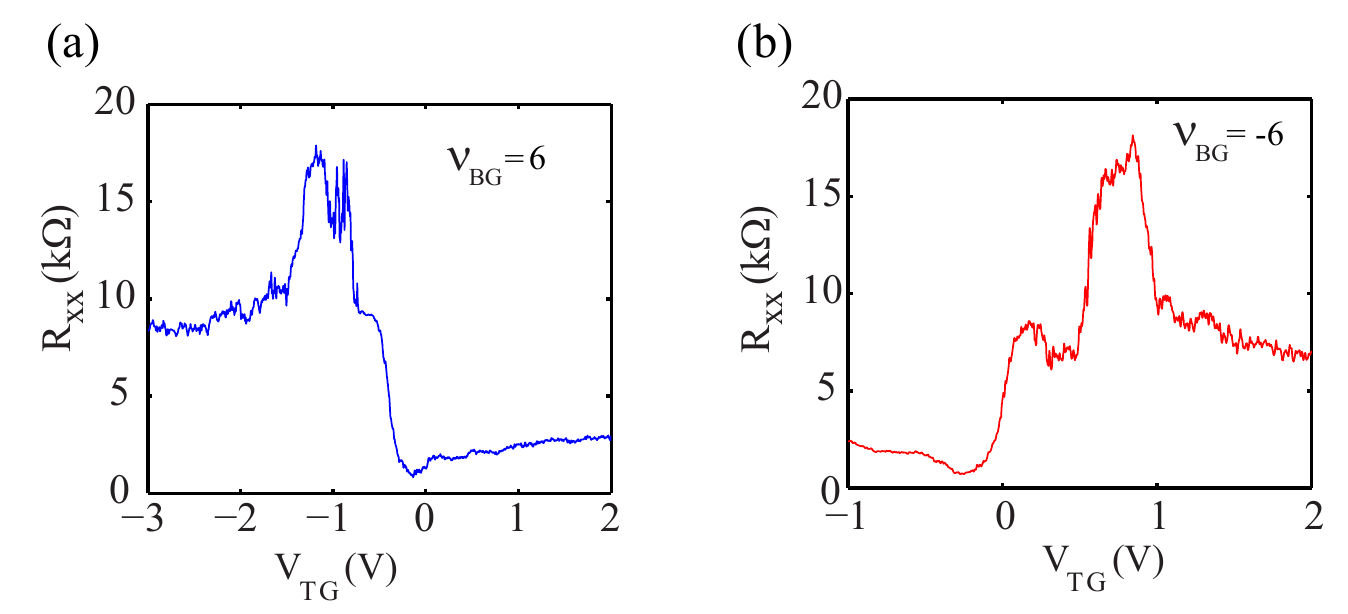}
		\caption{(color online) (a)(b) shows longitudinal resistance as a function of $V_{TG}$ for $\nu_{BG}=6,-6$ respectively.}
		\label{fig:images}
	\end{figure}
\end{center}

\section{Calculation of Diffusion constant}

The diffusion constant (D) is calculated using the expression \citep{bohra2012nonergodicity} 
$D=\frac{L_m v_F}{2} $, where $L_m$ is the mean free path and $v_F$ is the Fermi velocity.
The mean free path which is given by\cite{mayorov2011micrometer}
$L_m= \frac{h\mu}{2e} \sqrt[]{\frac{n}{\pi}}$ is 250 nm. We calculate the value of D to 
250 $\times 10^{-3} $ $m^2/s$ \\

\section{magneto resistance with bias}
\begin{center}
	\begin{figure}[h]
		\includegraphics[width=0.4\textwidth]{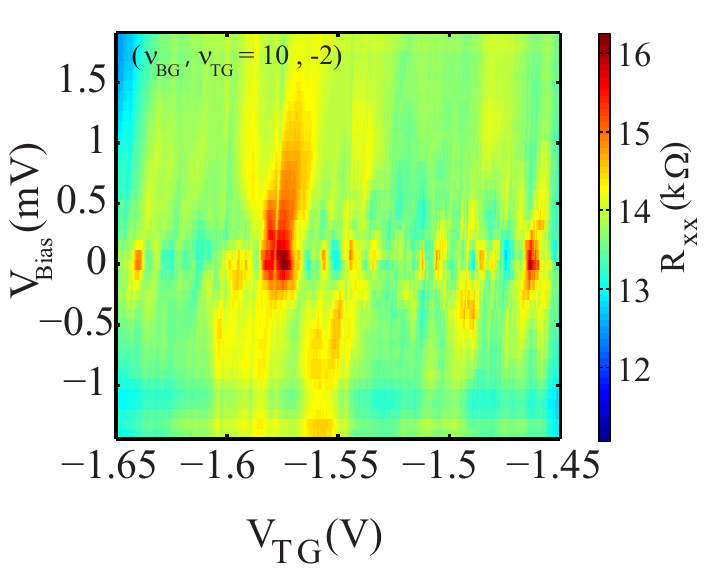}
		\caption{(color online) Figure shows $R_{xx}$ as a function of $V_{bias}$ and $V_{TG}$ for 10,-2 QH plateau}
		\label{fig:images}
	\end{figure}
\end{center}

Figure 8 shows that the longitudinal resistance ($R_{xx}$) as a function of bias for 10,-2 QH plateau. It can be clearly seen that the resistance fluctuation decreases with increasing bias voltage.
\section{$1/f$ noise measurement}
Low frequency $1/f$ noise is a versatile tool to study charge dynamics,disorder scattering, statics of defects and dielectric screening. For various semiconductor, nano wire and graphene $1/f$ noise follows the Hooge's emperical relationship:
$$\frac{S_{V}}{V^2}= \frac{\alpha_{H}}{f^{\beta}N}$$ where $\frac{S_{V}}{V^2}$ is the normalized voltage power spectral density,
$f$ is the frequency, $\alpha_{H}$ is called Hooge parameter and $N$ is the total number of charge carriers. The exponent $\beta$ is ideally expected to be close to 1.
The noise amplitude (A) is defined as: $A =$ $\alpha_{H}/N$.\\
\begin{center}
	\begin{figure}[h]
		\includegraphics[width=.64\textwidth]{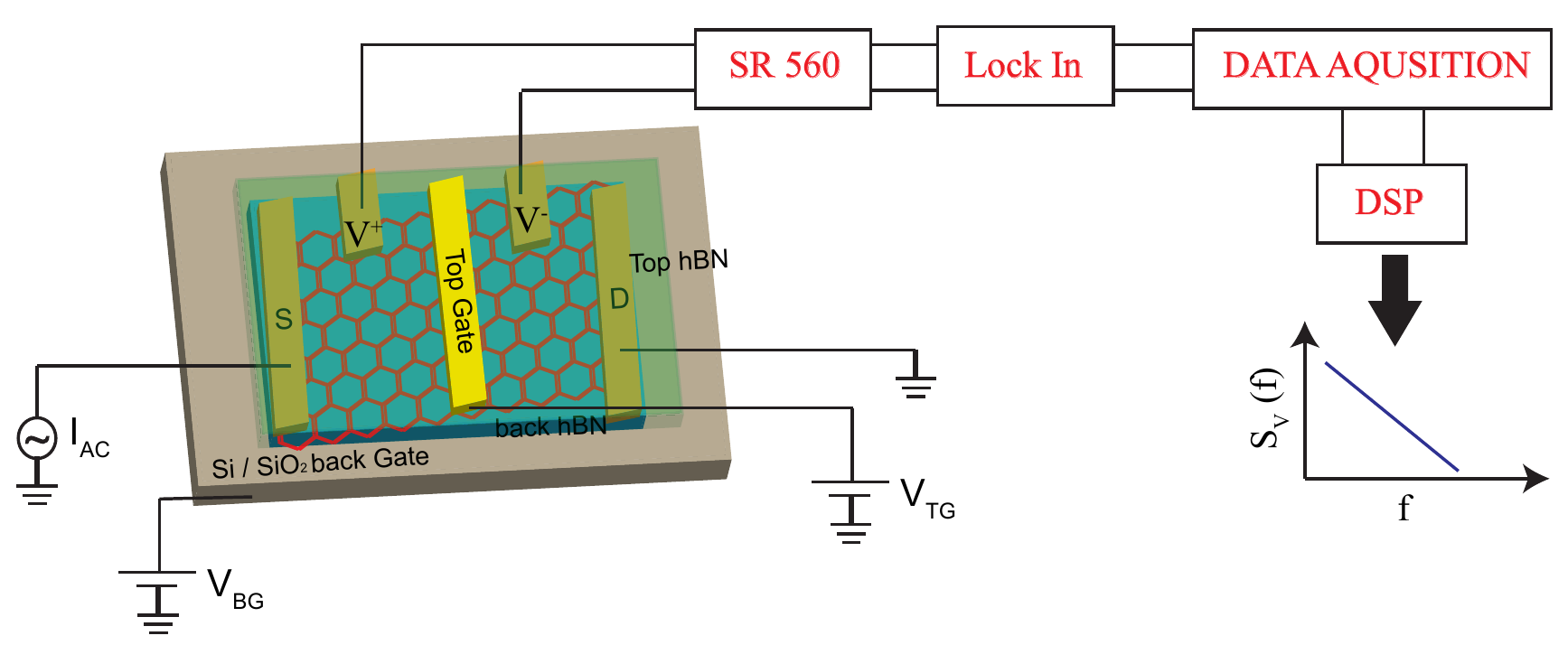}
		\caption{(color online) Schematic of $1/f$ noise measurement technique.}
		\label{fig:images}
	\end{figure}
\end{center}
The noise measurement was done in He3 cryostat at a base temperature of 240 mK using ac lock in technique.  The schematic of noise measurement technique is shown in figure 9. To measure noise, sample is biased with 25 nA current of frequency 228 Hz and the voltage fluctuation is measured with lock in amplifier using  high speed 16 bit digital to analog converter (NI 6210 DAQ) card. The output voltage is amplified using SR 560 voltage amplifier before it is fed to the input of lock in amplifier (SR 830). The data is taken for 15 minutes with lock in time constant of 1 ms with roll off of 24dB/octave. The data is sampled at a rate of 32768 Hz and decimation factor was kept at 128 which gives the effective sampling rate of 256 Hz. The final step is DSP (digital signal processing). In this step the acquired data is anti-aliased, down sampled and Welch's periodogram method is used for calulating PSD.
\section{Measurement performed on second device}

Figure 10 shows the result obtained on another single layer graphene device with
top gate length of 1.25 $\mu m$, width of 2.1 $\mu m$ and mobility of 9000 $cm^2/Vs$. 
Figure 10 (a) shows two probe conductance (in units if $\frac{e^2}{h}$) as a function of top gate voltage for different set of magnetic field at $V_{BG}=$ -28 V.  Figure 10(b) shows transconductance as a function of
magnetic field and top gate voltages for $V_{BG}$=-28 V. As can be seen from figure 10(a) and (b) at low magnetic field clear QH plaetau is seen in both  unipolar and bipolar regime but as the field increases we start getting well quantized plateaus in unipolar regime but fluctuations in bipolar regime. Figure 10(c) shows transconductance as a function of $V_{BG}$ and $V_{TG}$. In all of them (figure 4) we see clear Landau Level (LL) plateau in the unipolar region while in the bipolar regime we see fluctuation superimposed QH plateaus.

\begin{center}
	\begin{figure}[h]
		\includegraphics[width=1.07\textwidth]{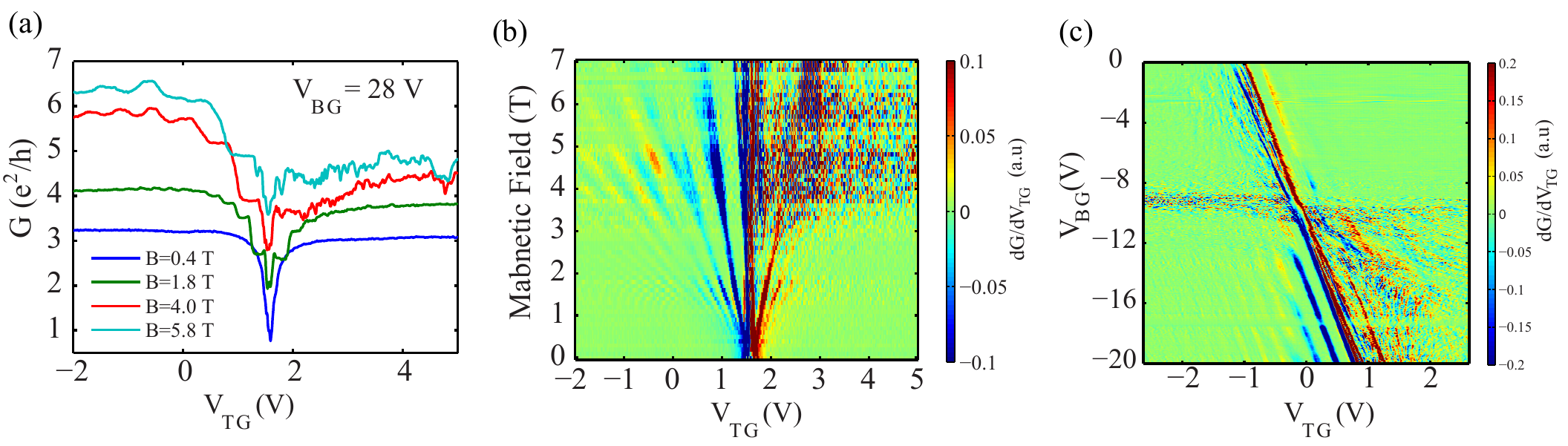}
		\caption{(color online) (a) shows resistance as a function of $V_{TG}$ for different set of magnetic field. Curve are shifted for clarity. (b) shows the Fan diagram at $V_{BG}=28 V$ (c) shows the resistance as a function of $V_{BG}$ and $V_{TG}$ at 6 T. All measurement are perforemd at 240 mK}
		\label{fig:images}
	\end{figure}
\end{center}
\bibliographystyle{apsrev4-1}
\bibliography{reference}

\end{document}